\documentclass[10pt]{elsarticle}

\pdfoutput=1

\usepackage[LGR,T1]{fontenc}
\usepackage[utf8]{inputenc}
\usepackage[english]{babel}
\usepackage{amsmath}
\usepackage{amssymb}
\usepackage{bm}
\usepackage{t1enc}
\usepackage{graphicx}
\usepackage{setspace}
\usepackage{lineno}

\newcommand{\rmd}{\mathrm d}

\begin{document}

\begin{frontmatter}

\title{Impact of gamma$'$ particle coarsening on the critical resolved shear stress of nickel-base superalloys with low aluminium and/or titanium content}

\author[PSI]{Péter Dusán Ispánovity\corref{cor1}}
\author[CEA]{Botond Bakó}
\author[KIT]{Daniel Weygand}
\author[PSI]{Wolfgang Hoffelner}
\author[PSI]{Maria Samaras}
\cortext[cor1]{Corresponding author. Tel.: +41 56 310 4563; fax: +41 56 310 4595. \emph{E-mail address:} peter.ispanovity@psi.ch (P.D. Ispánovity)}
\address[PSI]{Paul Scherrer Institut, CH-5232 Villigen PSI, Switzerland}
\address[CEA]{CEA, DEN, Service de Recherches de M\'etallurgie Physique, 91191 Gif sur Yvette, France}
\address[KIT]{Karlsruhe Institute of Technology, izbs, Kaiserstra{\ss}e 12, D-76131 Karlsruhe, Germany}

\begin{abstract}
In Ni-base superalloys with low Al and/or Ti content the precipitation and subsequent coarsening of $\gamma'$ particles at intermediate temperatures contribute to the degradation of the mechanical properties of the alloy. In the present paper the coarsening process is modelled and the change of the critical resolved shear stress of the alloy due to coarsening of the $\gamma'$ particles is calculated by means of statistical analysis of the depinning of a single gliding edge dislocation. It is found that the contribution of $\gamma'$ hardening to the critical resolved shear stress at 973 K reduces to more than half of its original value in less than one year.
\end{abstract}


\end{frontmatter}


\sloppy

\section{Introduction}

Based on their excellent high temperature strength and exceptional resistance to oxidation in a wide range of corrosive media, Ni-base superalloys are promising structural materials for next generation nuclear power plants. For example, a solid solution strengthened Ni-base superalloy (IN617) has been selected as the prime candidate for heat exchanger applications in Generation IV power plants \cite{geniv,Corwin2008}. These materials usually contain, among others, a few wt\% Al and/or Ti solutes. Upon heat treatment precipitation of a fine distribution of $\gamma'$ particles (Ni$_3$Al or Ni$_3$Ti) is observed with a coherent L1$_2$ crystal structure. These $\gamma'$ precipitates obstruct the motion of dislocations, resulting in high strength.

After exposure of these alloys to high temperatures for long times, however, the exceptional mechanical properties start to degrade. An important reason for this is the gradual coarsening of the $\gamma'$ particles. Since the volume fraction of the $\gamma'$ phase is nearly constant, the small particles disappear and the mean distance between the precipitates monotonously increases with exposure time. This implies that the critical resolved shear stress (CRSS) needed for a dislocation to overcome the obstacles decreases, resulting in the decrease of the material's strength. In the particular case of alloy IN617 (Table \ref{tab:comp}), which is used as a reference throughout the paper, in the temperature range between 873 to 973 K, the precipitation of spherical $\gamma'$ particles with mean diameter around 20 nm was observed for a volume fraction of $f \approx 2-4 \%$ \cite{Penkalla2001, Wu2008}. After a heat treatment at 973 K for 65$\,$600 hours the mean diameter increased to about 200 nm and in parallel, a drop in the microhardness was detected \cite{Wu2008}.

\setlength{\tabcolsep}{5pt}
\begin{table}[ht!]
\begin{center}
\begin{tabular}{lccccccccccccc}
\hline
Element & Ni & Cr & Co & Mo & Al & Fe & Mn & Ti & Si & C & Cu & B & S \\
\hline
Fraction (wt\%) & 52.2 & 22.0 & 12.5 & 9.0 & 1.2 & 1.0 & 1.0 & 0.4 & 0.1 & 0.1 & 0.5 & 0.006 & 0.015 \\
\hline
\end{tabular}
\caption{\label{tab:comp} Nominal chemical composition (wt\%) of the Inconel 617 alloy \cite{Wu2008}.}
\end{center}
\end{table}

The yield stress of a material is usually assumed to be the combination of different strengthening mechanisms (such as solute strengthening, work hardening, grain boundary strengthening, etc.). In this paper the focus is exclusively on the change in the flow stress due to coarsening of $\gamma'$ precipitates.

The paper is organized as follows: The $\gamma'$ coarsening modelled as a diffusion driven Ostwald ripening process is described in Section \ref{sec:ostwald}. A method based on volume diffusion through the matrix is applied. The discrete dislocation dynamics (DDD) model and the procedure to determine the CRSS of the alloy is presented in Section \ref{sec:results}. The simulation results are then summarized and the paper ends with concluding remarks.

\section{Simulation of $\gamma'$ coarsening}
\label{sec:ostwald}

In the applied model, the precipitates are considered to be hard spheres in a cubic box with periodic boundary conditions. To derive the equations of motion governing the coarsening the following assumptions are made: (i) the growth is isotropic, (ii) only volume diffusion is present, (iii) the diffusion is slow, hence the coarsening is considered quasistatic and (iv) the volume fraction of the precipitates is low (maximum few \%). The first assumption is usually not valid in $\gamma-\gamma'$ systems, since the precipitates tend to align in (100) directions, which leads to cubic or other anisotropic structures. At low volume fractions and small particles of $\gamma'$, however, spherical precipitates are observed experimentally. In the case of IN617, the $\gamma'$ particles are spherical even at the latest stages of annealing where the mean diameter is around 100-200 nm \cite{Penkalla2001, Wu2008}. This suggests, that the coarsening can be considered isotropic. The differential equations to be solved for the radius of the $i$th precipitate $R_i$ can be then written as \cite{Beenakker1986, Wang2004}
\begin{equation}
	\frac{\rmd R_i}{\rmd t} = - \frac{B_i}{R_i^2} \qquad (i=1,2,\dots,n),
\label{eqn:diff_eq}
\end{equation}
where $n$ is the total number of particles and the growth rates $B_i$ are calculated from the linear equations
\begin{equation}
	\bm A \cdot \bm B = \bm U,
\label{eqn:lin_eq}
\end{equation}
where $\bm A$ and $\bm U$ are $n \times n$ and $n \times 1$ matrices, respectively, and $\bm B = (B_1, \dots, B_n)$. The $\bm A$ matrix and the $\bm U$ vector can be constructed at every time step from the coordinates and radii of the precipitates -- for their actual form (taken from \cite{Wang2004}) see the appendix. We note that the variables in the above equations are adimensional \cite{Beenakker1986}.

There are two technical issues that need to be mentioned here. First, the simulations start with an initial precipitate number of several ten thousands. The complexity of the solution of Eq.~(\ref{eqn:lin_eq}) is $\mathcal{O}(n^3)$, and has to be solved at every time step, which is computationally too expensive. The complexity can be reduced, however, because there is a screening length $\lambda \approx \langle R \rangle/f^{1/2}$ for diffusion, above which, in principle, no interactions take place between the precipitates \cite{Beenakker1986} (where $\langle R \rangle$ stands for the mean precipitate radius). As such, the simulation cube can be split up into distinct regions, and several equations can be solved with lower dimensions simultaneously. The applied method is similar to that used by Beenakker \cite{Beenakker1986}.

The second technical issue is related to the numerical solution of Eq.~(\ref{eqn:diff_eq}). For the integration we use the 4th order Runge-Kutta scheme with adaptive step size. The latter is necessary, since there are slow and fast varying regions during the solution. To demonstrate this, consider the event of the disappearance of a small precipitate with index $i$, i.e.\ $R_i \to 0$. In this case in the $i$th equation of Eq.~(\ref{eqn:lin_eq}), according to (\ref{eqn:matrix_def}) the diagonal element becomes the dominating term in $\bm A$, and the $i$th equation tends to $(1/R_i) B_i = 1/R_i$. Thus, in this case $B_i \to 1$ and the derivative of $R_i$ in Eq.~(\ref{eqn:diff_eq}) diverges. To sum up, at every disappearance of a particle, very small time steps are required, while between such events longer time steps are acceptable.

In practice, the adaptive time step size still does not provide optimal performance. Due to the diverging derivatives described above, extremely small timesteps are chosen, but these small particles usually have negligible volume ($V_i \propto R_i^3$) and thus they hardly affect the coarsening kinetics of the other particles. To solve this problem, Eq.~(\ref{eqn:lin_eq}) was slightly modified by taking $\bm U'$ on the right hand side as
\begin{equation}
 U'_i := \frac{f(R_i)}{R_i} - \frac{\langle R \rangle_f}{\langle R \rangle}
\end{equation}
for every $i$, where for the arbitrary $f$ function the following properties hold:
\begin{enumerate}
	\item $f(x) > 0$ if $x>0$,
	\item $f(x) \to 0$ exponentially as $x \to 0$,
	\item $f(x) \to 1$ exponentially as $x/R_\text{min} \gg 1 $, with $R_\text{min}$ being a characteristic minimum radius for the precipitates,
\end{enumerate}
and $\langle R \rangle_f := (1/n) \sum_i f(R_i)$. In the present case $f(x) := [1-\exp(-x/R_\text{min})]^2$ was chosen. It is easy to see that if all the radii $R_i \gg R_\text{min}$, then there is no detectable difference in the equations, and when for an $i$ $R_i \to 0$, then $B_i \to 0$ exponentially, and the derivatives in Eq.~(\ref{eqn:diff_eq}) no longer diverge. In addition, the requirement of the constancy of the total volume of the precipitates, which is expressed by $\sum_i B_i = 0$ \cite{Beenakker1986}, is found not to be violated. To test the validity of the outlined approximation simulations were run with different $R_\text{min}$ values as well as without this modification, and it was found that (at low enough $R_\text{min}$ values) the introduced error is negligibly small. In these calculations, the lattice constant has been taken as the value of $R_\text{min}$.

\begin{figure}[!ht]
\begin{center}
\includegraphics[angle=0, width=4cm]{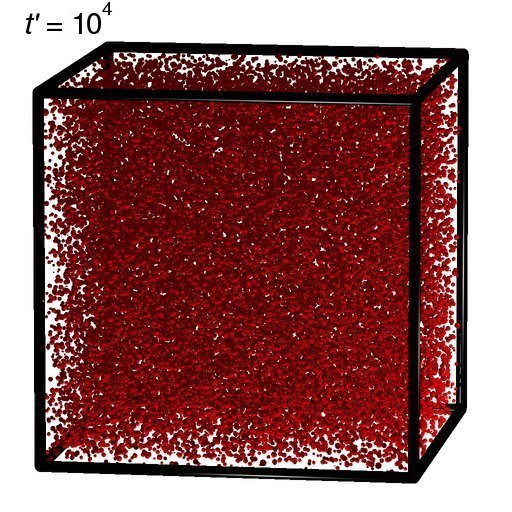}
\includegraphics[angle=0, width=4cm]{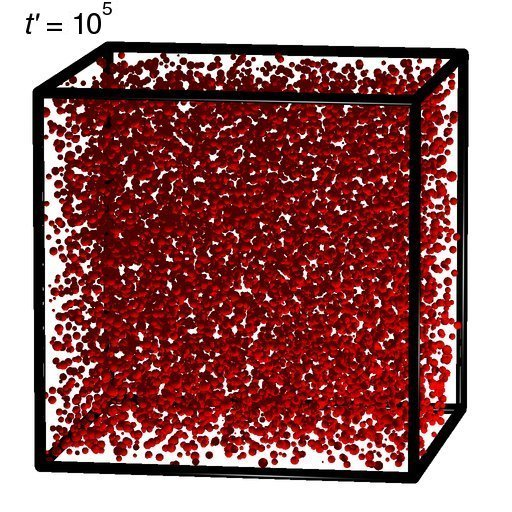}
\includegraphics[angle=0, width=4cm]{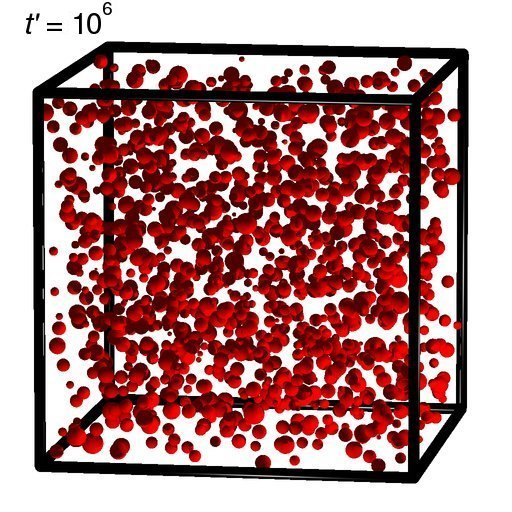}
\caption{\label{fig:coarsening} Coarsening of the $\gamma'$ particles in the $L = 1.75$ $\mu$m edge box at $f=4\%$ volume content. The mean radius $\langle R \rangle$ increases from 10 nm on the left image to 33 nm on the right.}
\end{center}
\end{figure}

The simulations were performed with a volume fraction of $f = 4\%$ and with three different cube sizes: $L = 1.4$ $\mu$m, $L = 1.75$ $\mu$m and $L = 2.25$ $\mu$m. Initially, the particles were placed at random positions with a lognormal radius distribution, with initial mean radius of $R_0 = 10$ nm and relative standard deviation of 20\%. Figure \ref{fig:coarsening} shows an example sequence for $L = 1.75$ $\mu$m. Note that the initial number of particles was around 20$\,$000 for $L=1.4$ $\mu$m 50$\,$000 for $L=1.75$ $\mu$m and 100$\,$000 for $L = 2.25$ $\mu$m.

The coarsening process has been the subject of extensive theoretical studies. For low ($\lesssim 10 \%$) volume fractions the Lifsitz-Slyozov-Wagner (LSW) theory provides an exact solution for the asymptotic (long time) behaviour of the system \cite{Lifshitz1961,Wagner1961}. In particular, the mean particle radius $\langle R \rangle$ increases as
\begin{equation}
	\langle R \rangle(t)^3 - R_0^3 = k t
\label{eqn:R_time_dep}
\end{equation}
at large enough times, where $R_0$ is the mean radius at $t=0$ and $k$ is a constant depending on the diffusion coefficient, the volume fraction and possibly on the initial spatial distribution of the particles. Since there is no available data for diffusion constants in the studied temperature range and chemical composition, estimating the coarsening coefficient $k$ relies on derivation from experimental observations. Namely, determining $\langle R \rangle$ is experimentally possible (for example using TEM images), from which $k_\text{exp}$ can be obtained by fitting Eq.~(\ref{eqn:R_time_dep}) on such experimental data. This was performed on several Ni-base superalloys in the past, mainly with an Al/Ti content in the range of $6-10$ wt.\% \cite{Ardell1966,Ardell1970,Chellman1974,Irisarri1985}, and all authors confirmed the coarsening kinetics of Eq.~(\ref{eqn:R_time_dep}). More recently, a material with lower $\gamma'$ volume fraction was investigated by Zhao and co-workers \cite{Zhao2004}. In that case the alloy contained among others, 0.75 wt.\% Al and 1.58 wt.\% Ti and a similar coarsening behaviour was observed as in the case of IN617 \cite{Wu2008}. The measured $k_\text{exp}$ followed
\begin{equation}
	k_\text{exp}(T) = k_0 \exp\left(-\frac{Q}{\frac12 RT}\right)
\label{eqn:k_exp}
\end{equation}
with an activation energy of $Q=247$ kJ mol$^{-1}$ and $k_0=1.4\times 10^{27}$ nm$^3\cdot$h$^{-1}$ \cite{Zhao2004}. The coarsening rate coefficient $k$ has not been measured for IN617 (to the authors' knowledge), and as such it is assumed that it does not  differ too much from the value given by Eq.~(\ref{eqn:k_exp}).

\begin{figure}[!ht]
\begin{center}
\hspace*{-0.4cm}
\includegraphics[angle=0, width=6.5cm]{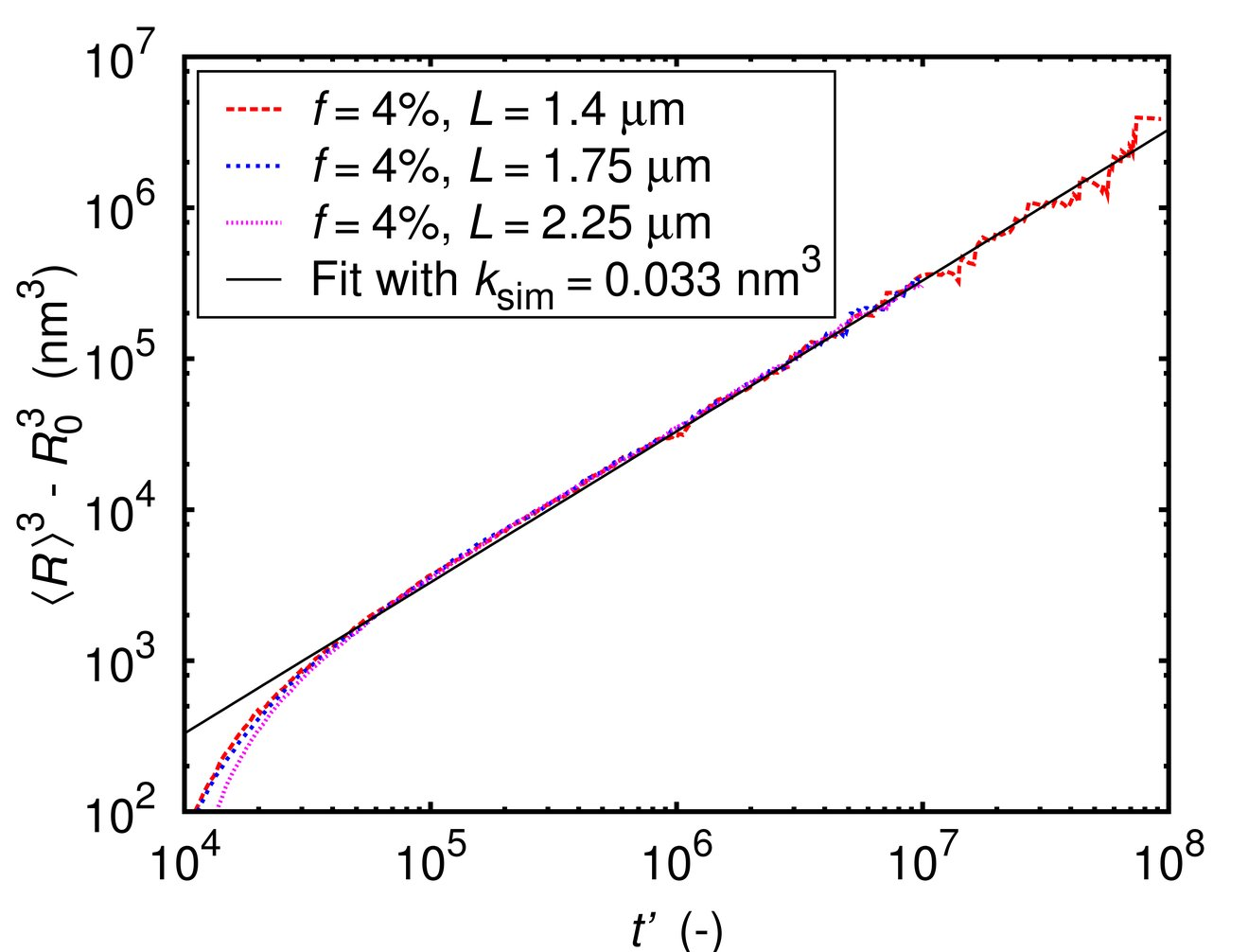}
\begin{picture}(0,0)
\put(-196,125){\normalsize\sffamily{(a)}}
\end{picture}
\hspace*{0.5cm}
\includegraphics[angle=0, width=6.5cm]{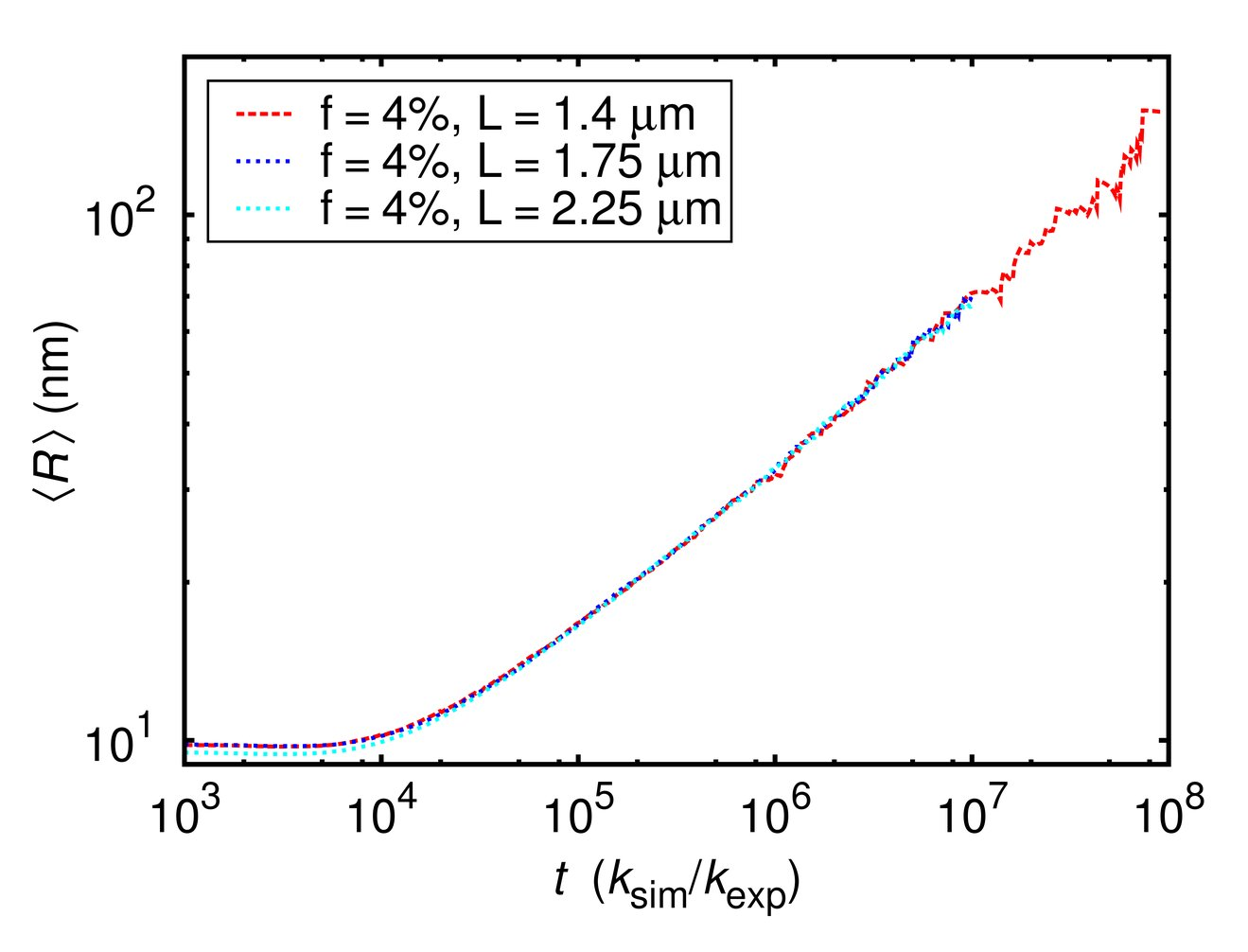}
\begin{picture}(0,0)
\put(-190,125){\normalsize\sffamily{(b)}}
\end{picture}
\caption{\label{fig:R_time_dep} (a) The time dependence of the cube mean radius $\langle R \rangle^3$ at different simulation parameters. The fitted solid line corresponds to Eq.~(\ref{eqn:R_time_dep}). (b) The evolution of the mean radius $\langle R \rangle(t)$.}
\end{center}
\end{figure}
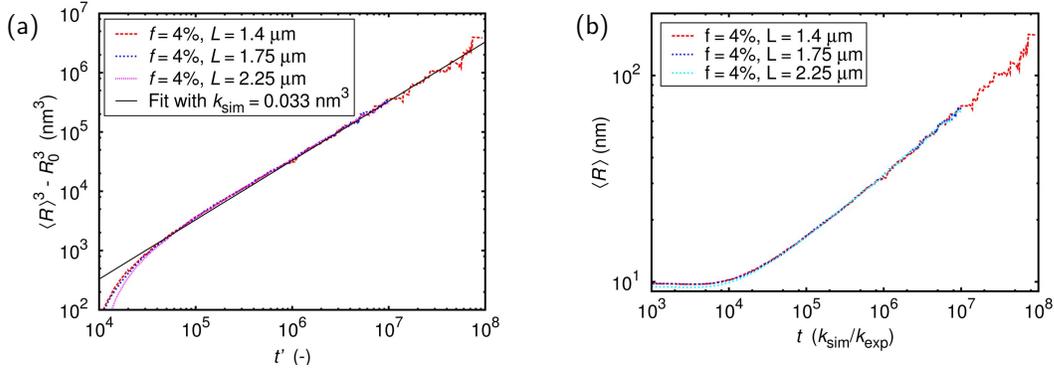

It was already mentioned above, that in the simulations dimensionless time units are used. However, in order to give a life time assessment for the considered materials, it is necessary to link the coarsening process to a real time scale. This can be done using Eq.~(\ref{eqn:R_time_dep}) again, which is, according to Fig.~\ref{fig:R_time_dep}(a), valid in the simulations too. The adimensional simulation time $t'$ can be transformed into real time $t$ with a linear transformation $t = (k_\text{sim}/k_\text{exp}) t'$, where $k_\text{sim}$ is the fitted growth rate in simulations, and $k_\text{exp}$ is the one measured in the experiments. As such, in Figs.~\ref{fig:coarsening} and \ref{fig:R_time_dep}, the simulation time units are displayed as $(k_\text{exp}/k_\text{sim})t$. The linear fit of Eq.~(\ref{eqn:R_time_dep}) in Fig.~\ref{fig:R_time_dep}(a) yields
\begin{equation}
	k_\text{sim} = 3.3 \times 10^{-2} \text{ nm}^3.
\end{equation}
Figure \ref{fig:R_time_dep}(b) shows the evolution of the mean radius $\langle R \rangle(t)$ itself with the same time units. It can clearly be seen that the finite size effects are negligible at these simulation box sizes.

\section{Results}
\label{sec:results}

\subsection{Simulation set up}
\label{sec:crss}

To determine the CRSS of the alloy the threshold shear stress needed for a single edge dislocation to continuously penetrate through the $\gamma'$ precipitates was calculated. Such investigations have already been performed several authors on similar materials to the one considered here \cite{Mohles2001, Mohles2002, Mohles2004, BakoWeygand2007, Bako2008, Bako2009, BakoSamaras2009}. The main difference is that in the present study, instead of adopting a certain size and spatial distribution for the particle arrangement, the data has been obtained from the direct simulation of the Ostwald ripening process, described in the previous section. In this work, a statistical method \cite{Bako2008} is used to obtain the CRSS. Its main features can be summarized as follows. The CRSS measured in a single simulation is a probabilistic variable. To obtain the CRSS, which is well defined only for systems of infinite size, many smaller systems with different realizations of the random precipitate distribution are considered instead of studying a system as large as possible. To determine the critical resolved shear stress of an infinite system, the size-dependent transition probability $P_L$ is defined as the fraction of realizations where, for a given shear stress and system size $L$ the dislocation can traverse the simulation box of size $L$. Transition probabilities $P_L$ for different system sizes $L$ are well approximated by
\begin{equation}
	P_L(\tau) = \frac{1}{2}\left[ 1 + \text{erf}\left( \frac{(\tau - \tau_\text{c}) L}{\sqrt{2} \sigma} \right) \right],
\label{eqn:cdf}
\end{equation}
where $L$ is the system size, $\sigma$ is a fitting parameter for the width of the distribution and $\tau_\text{c}$ is the critical stress. The form of the distribution function implies, that in the limit of $L\to \infty$ the function tends to a step-function at $\tau_\text{c}$, therefore the fitted $\tau_\text{c}$ can be considered to be the bulk CRSS \cite{Bako2008}.

\begin{figure}[!ht]
\begin{center}
\includegraphics[angle=0, width=5cm]{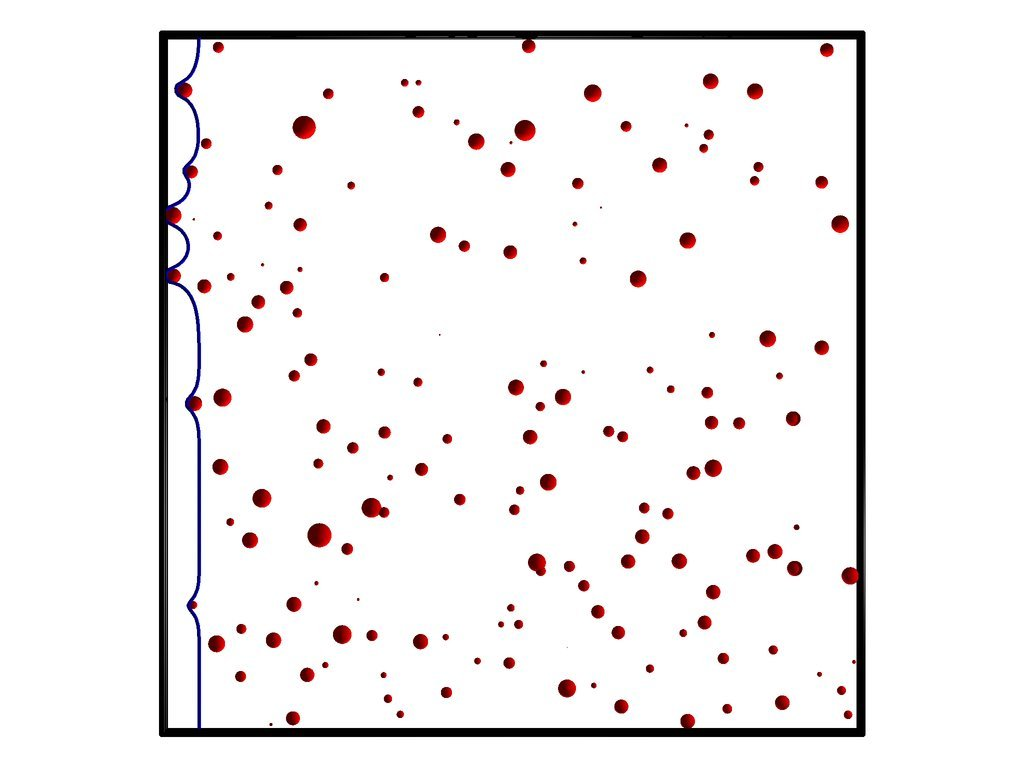}
\begin{picture}(0,0)
\put(-145,92){\normalsize\sffamily{(a)}}
\end{picture}
\hspace*{-0.6cm}
\includegraphics[angle=0, width=5cm]{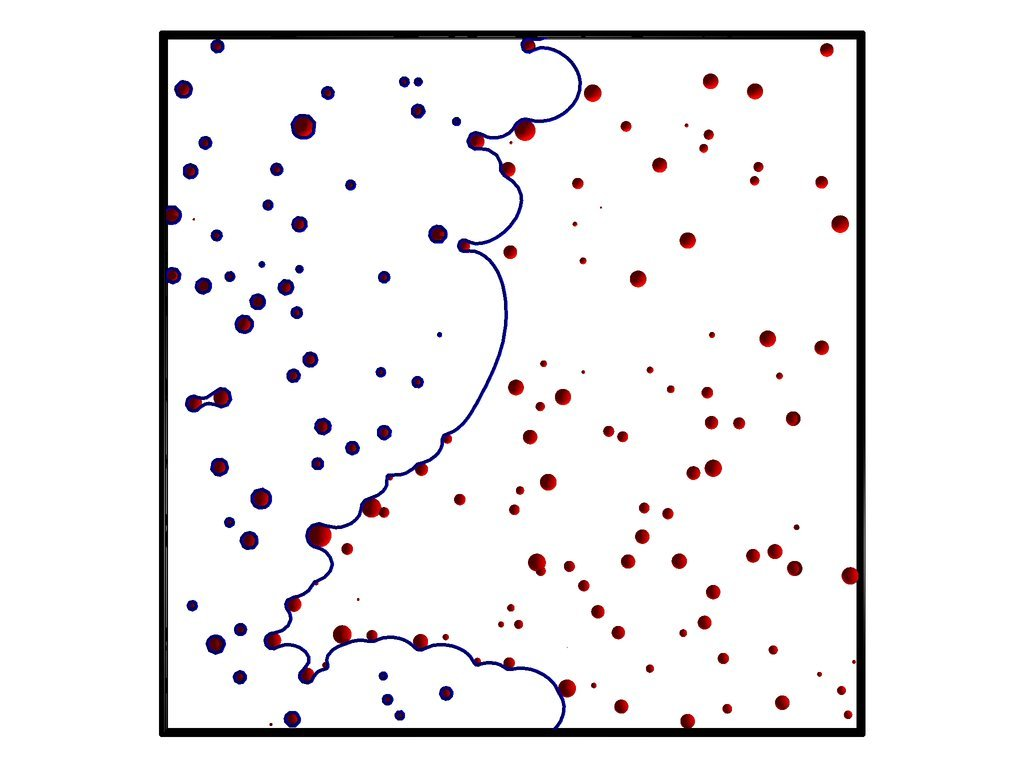}
\hspace*{-0.5cm}
\includegraphics[angle=0, width=5cm]{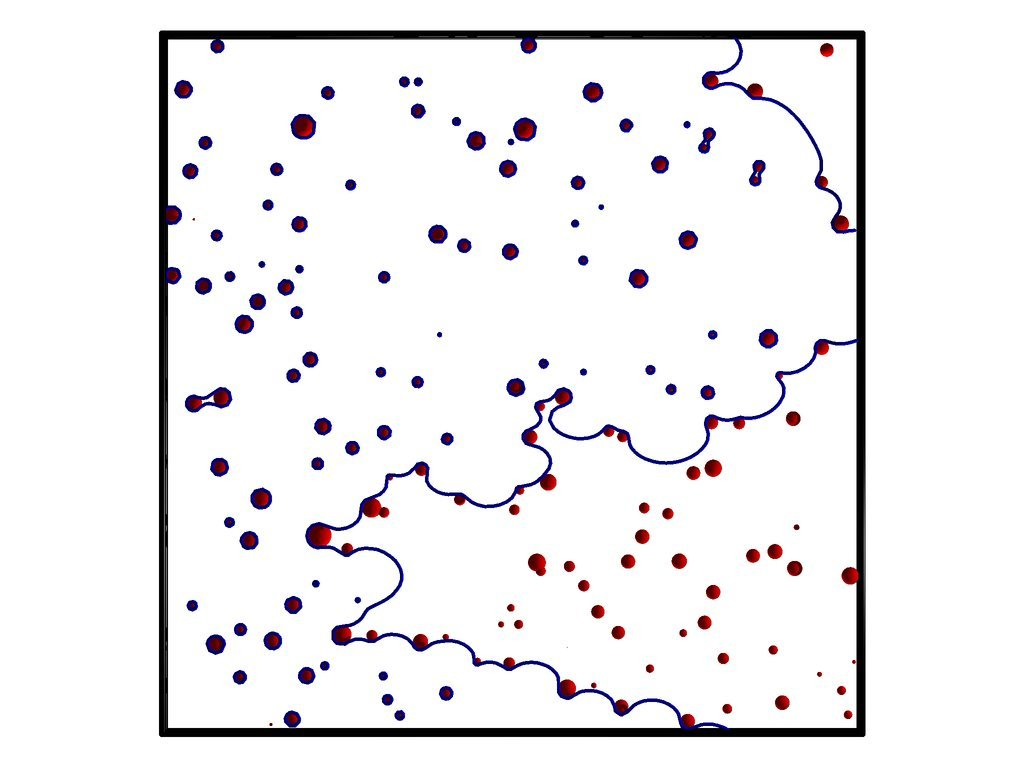}\\
\includegraphics[angle=0, width=5cm]{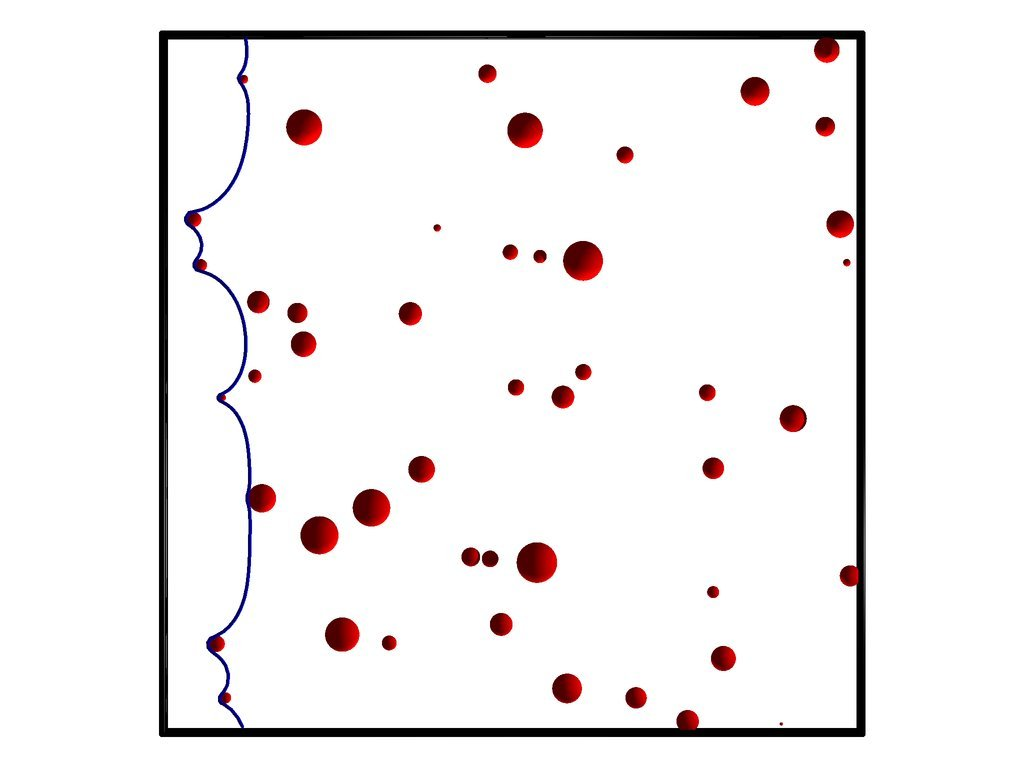}
\begin{picture}(0,0)
\put(-145,92){\normalsize\sffamily{(b)}}
\end{picture}
\hspace*{-0.6cm}
\includegraphics[angle=0, width=5cm]{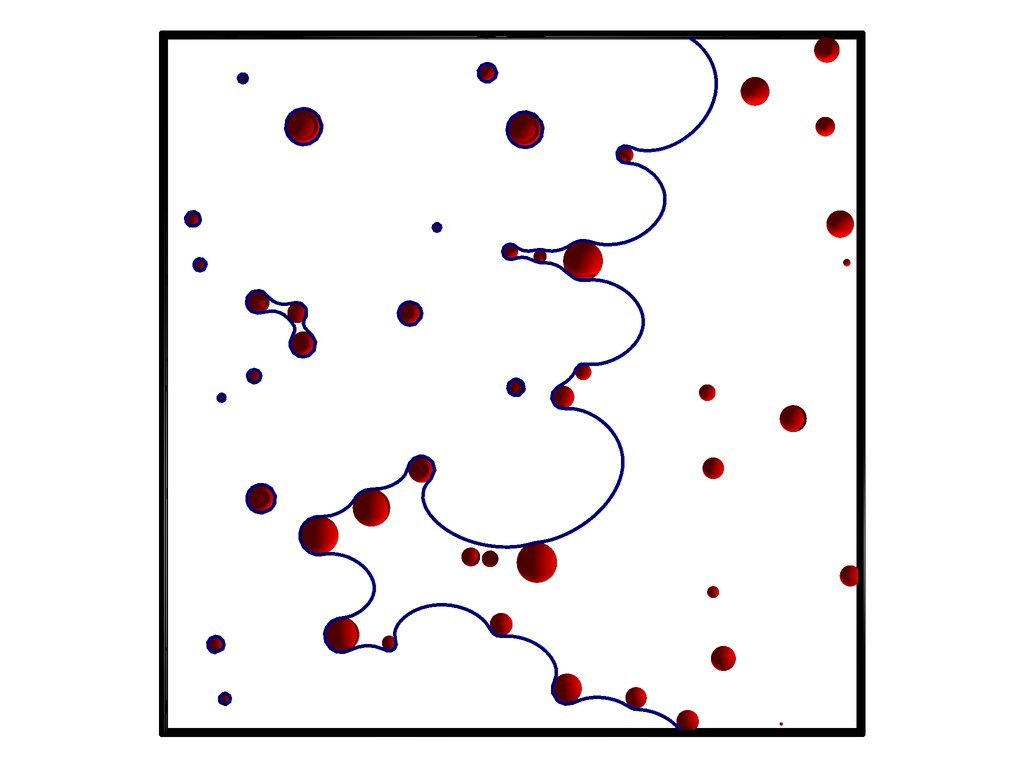}
\hspace*{-0.5cm}
\includegraphics[angle=0, width=5cm]{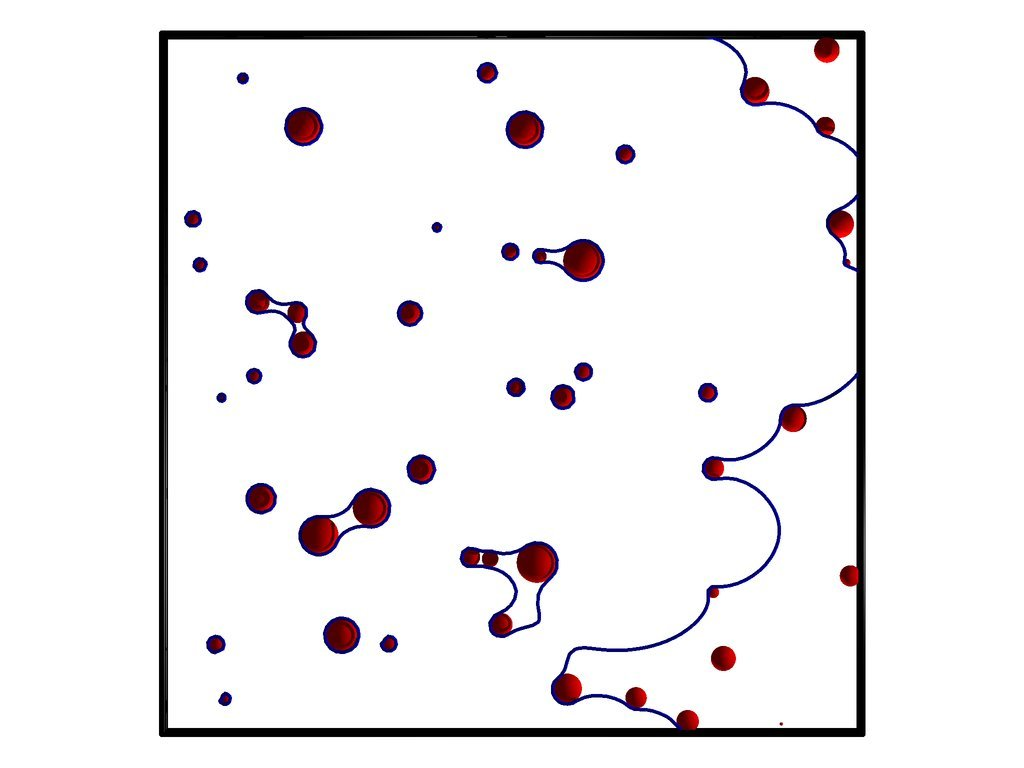}\\
\caption{\label{fig:movie_1} DDD simulation of an edge dislocation moving through the particles at constant applied stress and $L=1.75$ $\mu$m. Only the $\gamma'$ particles intersecting with the glide plane are shown. Sequence (a): $t=10^5 \, k_\text{exp}/k_\text{sim}$, sequence (b): $t=10^6 \, k_\text{exp}/k_\text{sim}$.}
\end{center}
\end{figure}
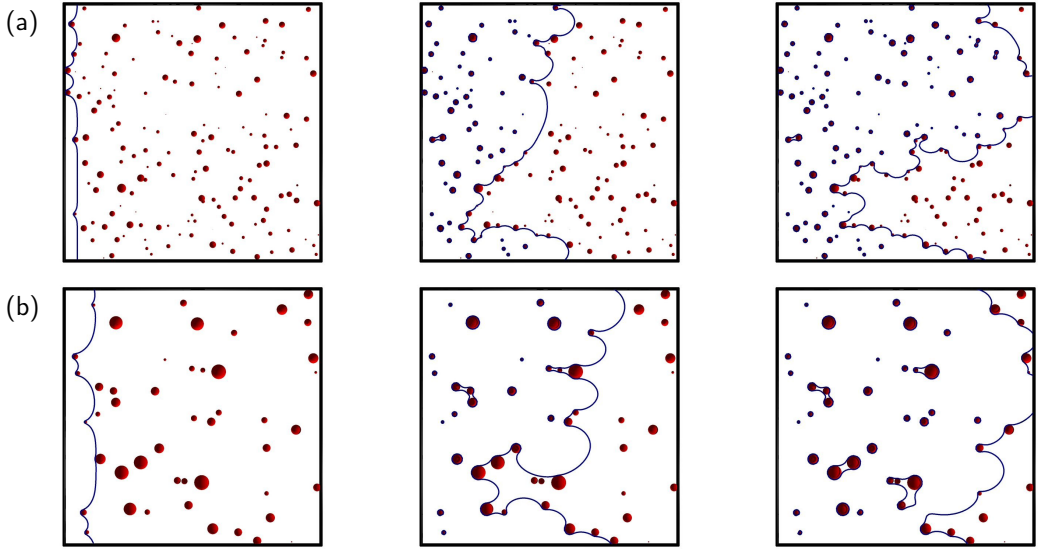

In the present case, slices were cut from the precipitate configurations of Fig.~\ref{fig:coarsening} with appropriate plane distances corresponding to the $(111)$ glide planes of the dislocations. The planes were parallel to one side of the simulation cube. Then DDD simulations of a single edge dislocation moving in its glide plane were performed, as discussed in \cite{BakoWeygand2007} (for the fundamentals of the DDD method see \cite{Weygand2002}). For the matrix we used \emph{fcc} crystal structure and the following material parameters (corresponding to room temperature IN617 \cite{In617_data}): lattice constant of $a=3.66$ \AA{}, shear modulus of $\mu = 81$ GPa and Poisson's ratio of $\nu = 0.3$. Two example simulation sequences are shown in Fig.~\ref{fig:movie_1} corresponding to two different times.

It has to be noted, that while the coarsening process is simulated in the temperature range of 873$-$1073 K, the CRSS determination is performed at room temperature by averaging over 90 realizations. This means that only dislocation glide is assumed, and temperature induced processes, such as dislocation climb or cross slip, are neglected.

In Ni-base superalloys, the transition temperature between the Orowan process and other high temperature bypassing mechanisms is around 1073 K \cite{Guo2007}. Below this value the relative drop in the yield stress can be attributed to the same relative drop in the elastic constants, which suggests that the bypassing mechanism is not changed. Our DDD model is, therefore, valid also in the temperature range of 873$-$1073 K, only the elastic constants should be modified. To extend these studies to even higher temperatures the inclusion of dislocation climb and cross slip into the model is the subject of further work.

\subsection{Determination of the CRSS}
\label{sec:discussion}

\begin{figure}[!ht]
\begin{center}
\hspace*{-0.4cm}
\includegraphics[angle=0, width=6.5cm]{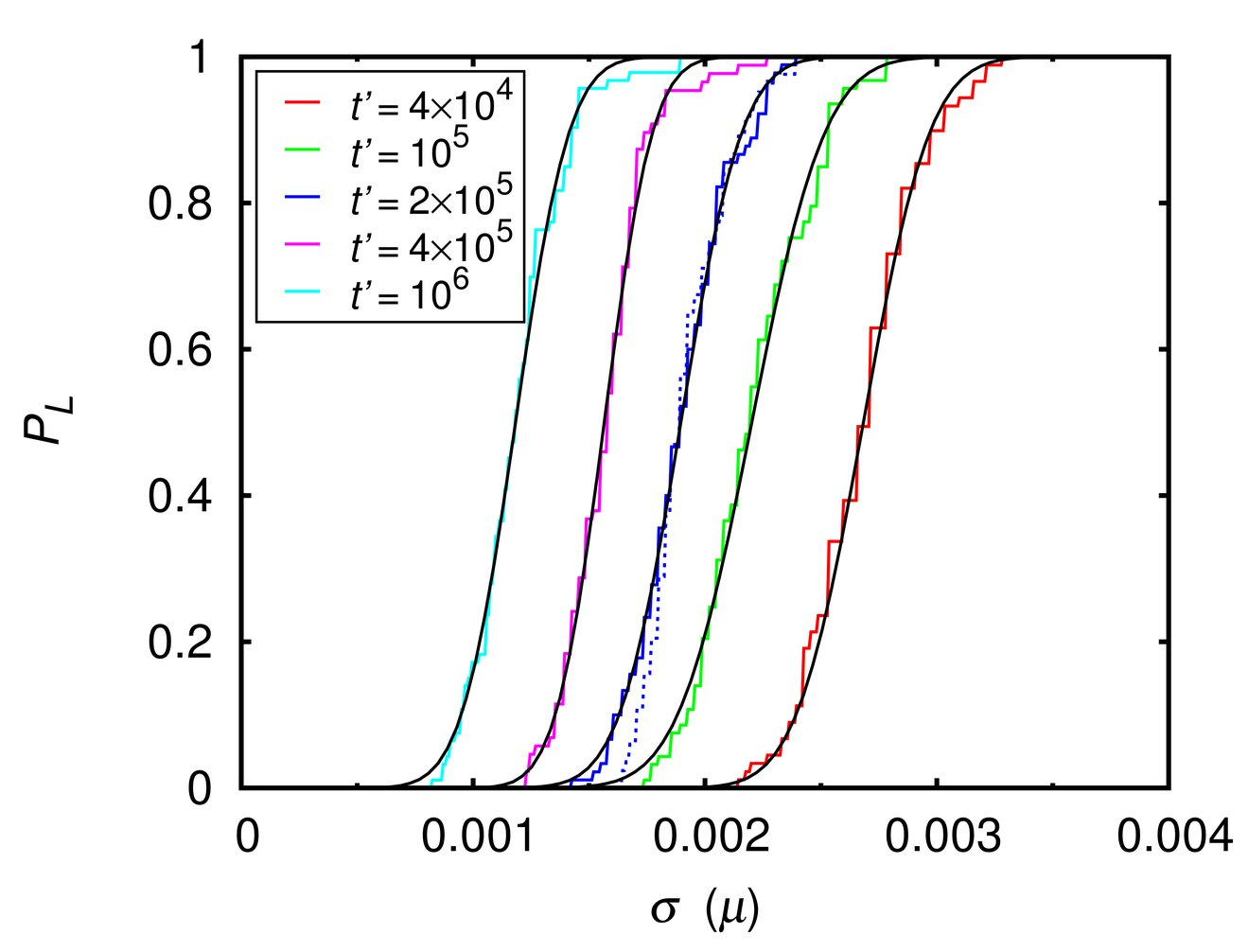}
\caption{\label{fig:cumul_distrib} The cumulative distribution function of the critical shear stress values. The solid and dashed colour lines correspond to a system size of $L=1.75$ $\mu$m and $L=2.25$ $\mu$m, respectively. The thin black lines are the fitted functions $P_L$ of Eq.~(\ref{eqn:cdf}) for $L=1.75$ $\mu$m.}
\end{center}
\end{figure}

The measured cumulative distribution functions of the CRSS of individual 2D slices for the $L=1.75$ $\mu$m sample are plotted in Fig.~\ref{fig:cumul_distrib} at different times. Each of the curves is constructed from $\sim90$ simulations.

\begin{figure}[!ht]
\begin{center}
\hspace*{0.0cm}
\includegraphics[angle=0, width=7.5cm]{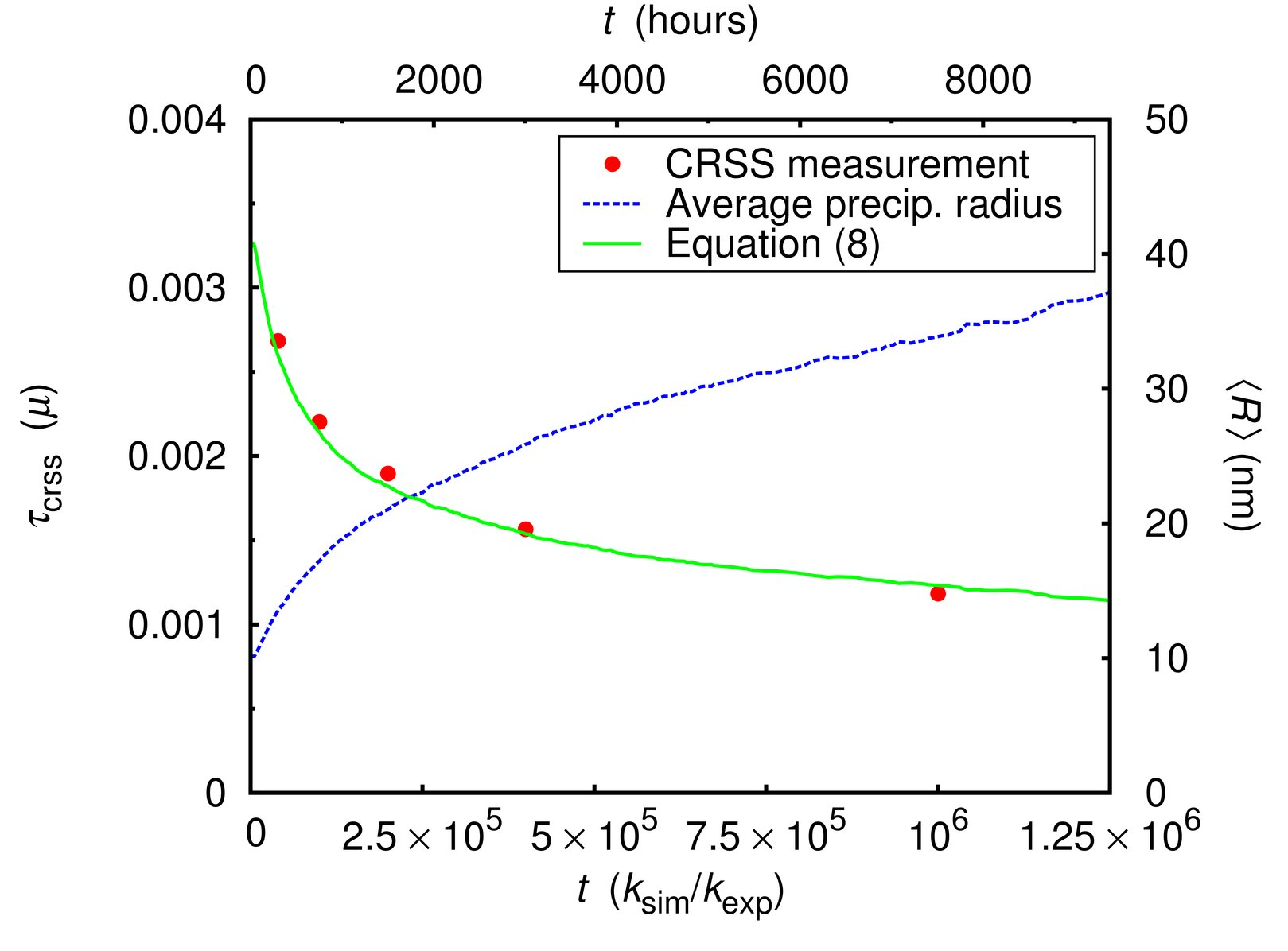}
\caption{\label{fig:crss} The evolution of the CRSS. The upper timescale assumes $T=973$ K. For comparison, the $\langle R \rangle(t)$ function was plotted again, as well as the theoretical expression of Eq.~(\ref{eqn:crss}) (green line) for the CRSS.}
\end{center}
\end{figure}

The determined CRSS as a function of time is shown in Figure \ref{fig:crss}. In the measured range the CRSS drops more than 50\%, while the mean radius of the particles increases from 10 nm to about 35 nm. To relate the dimensionless timescale to the real one, Eq.~(\ref{eqn:k_exp}) is implemented with $T = 973$ K. In this case $k_\text{exp} = 4.22$ nm$^3 \cdot$h$^{-1}$, and $k_\text{sim}/k_\text{exp} = 7.82\times 10^{-3}$ h. The top scale of Fig.~\ref{fig:crss} corresponds to this transformation, and shows that the drop of more than 50\% in the CRSS due to coarsening of precipitates happens in less than one year, even at this moderate temperature. For a more precise estimation for IN617, the experimental measurements of the coefficient $k_\text{exp}$ would be necessary.

Several analytical expressions exist for the CRSS due to a distribution of precipitates bypassed by the Orowan mechanism \cite{Nembach1996}. These are based on assumptions on the spatial and radial distribution of the particles. In our case, the coarsening starts from spatially random initial configurations with a lognormal size distribution (see Section \ref{sec:ostwald}). During the Ostwald ripening, both spatial correlations of the particles evolve and the size distribution also tends to an asymptotic function, described within the LSW theory \cite{Lifshitz1961,Wagner1961}. To estimate the effect of these phenomena on the CRSS our results are compared with the formula derived by Nembach based on the work of Bacon and co-workers \cite{Bacon1973,Scattergood1975}\footnote{The equation corresponds to the CRSS of an edge dislocation (the same case as in our simulations). In the case of screws, an additional $1/(1-\nu)$ factor has to be considered.}:
\begin{equation}
	\tau_\text{p}(\langle R \rangle) = \frac{\mu}{4 \pi} \frac{2b}{\omega_\text{L}\langle R \rangle} \ln(2 \omega_\text{H}\langle R \rangle/b).
\label{eqn:crss}
\end{equation}
In the formula $\omega_\text{L}$ and $\omega_\text{H}$ are numeric constants depending on the geometric properties of the precipitates. Here the values of $\omega_\text{L} = 5.69$ and $\omega_\text{H} = 1.22$ were adopted, which were derived for monodisperse particles \cite{Nembach1996}. In Fig.~\ref{fig:crss} the solid line corresponds to Eq.~(\ref{eqn:crss}), without any fitting parameters. The remarkably good agreement suggests that even if the spatial correlations affect the CRSS, the dependence must be weak or only effective in case of particle clustering and particle free areas.

\section{Conclusions}
\label{sec:conclusion}

A new method was established to study the effect of $\gamma'$ coarsening on the CRSS of Ni-base superalloys with a low volume fraction of $\gamma'$ phase. The assumptions taken in this model are that the coarsening is driven by bulk diffusion through the matrix (Ostwald ripening), and that the volume fraction of the precipitates remains constant during the growth. The timescale of the growth process was determined by fitting to experimental observations. The critical resolved shear stress of the material was measured at different stages of the coarsening with a DDD method coupled with a statistical analysis. It was found that the spatial inhomogeneities introduced during the coarsening do not have significant effect on the CRSS and that in an IN617 alloy at 973 K the CRSS contribution of the $\gamma'$ phase decreased by more than 50\% of its initial value in less than 1 year.

\section*{Appendix}
\appendix

The form of the matrices introduced in Eq.~(\ref{eqn:lin_eq}) \cite{Wang2004}:
\begin{equation}
\bm A =
\begin{pmatrix}
	\frac{1}{R_1} - \frac{1}{n\langle R \rangle} \sum\limits_{j\ne 1}^n \frac{R_j}{r_{j1}} & \frac{1}{r_{12}} - \frac{1}{n\langle R \rangle} \sum\limits_{j\ne 2}^n \frac{R_j}{r_{j2}} & \dots & \frac{1}{r_{1n}} - \frac{1}{n\langle R \rangle} \sum\limits_{j\ne n}^n \frac{R_j}{r_{jn}} \\
	\frac{1}{r_{21}} - \frac{1}{n\langle R \rangle} \sum\limits_{j\ne 1}^n \frac{R_j}{r_{j1}} &	\frac{1}{R_2} - \frac{1}{n\langle R \rangle} \sum\limits_{j\ne 2}^n \frac{R_j}{r_{j2}} & \dots & \frac{1}{r_{2n}} - \frac{1}{n\langle R \rangle} \sum\limits_{j\ne n}^n \frac{R_j}{r_{jn}} \\
	\vdots & \vdots & \ddots & \vdots \\
	\frac{1}{r_{n1}} - \frac{1}{n\langle R \rangle} \sum\limits_{j\ne 1}^n \frac{R_j}{r_{j1}} & \frac{1}{r_{n2}} - \frac{1}{n\langle R \rangle} \sum\limits_{j\ne 2}^n \frac{R_j}{r_{j2}} & \dots & \frac{1}{R_n} - \frac{1}{n\langle R \rangle} \sum\limits_{j\ne n}^n \frac{R_j}{r_{jn}}
	
\end{pmatrix},
\quad
\bm U =
\begin{pmatrix}
	\frac{1}{R_1} - \frac{1}{\langle R \rangle} \\
	\frac{1}{R_2} - \frac{1}{\langle R \rangle} \\
	\vdots \\
	\frac{1}{R_n} - \frac{1}{\langle R \rangle}
\end{pmatrix},
\label{eqn:matrix_def}
\tag{A1}
\end{equation}
where $r_{ij}$ denotes the distance between the $i$th and $j$th particles.

\bibliographystyle{model1a-num-names}
\bibliography{../../bibtex/journalss.bib,../../bibtex/database.bib}

\end{document}